\documentclass[a4paper,fleqn,usenatbib]{mnras}
\usepackage{amsmath}	
\usepackage[T1]{fontenc}
\usepackage{ae,aecompl}

\usepackage{graphicx}	

\title[Dynamical friction on hot bodies]{Dynamical friction on hot
  bodies in opaque, gaseous media}

\author[F. Masset \& D. Velasco]{
Fr\'ed\'eric S. Masset \thanks{E-mail: masset@icf.unam.mx}
and David A. Velasco Romero
\\
Instituto de Ciencias F\'\i sicas, Universidad Nacional Aut\'onoma de
M\'exico, Av. Universidad s/n, 62210 Cuernavaca, Mor., Mexico
}

\date{Accepted Nov. 17, 2016;  Received Nov. 1, 2016; in original form
  Sep. 10, 2016}

\pubyear{2016}

\begin{document}
\label{firstpage}
\pagerange{\pageref{firstpage}--\pageref{lastpage}}
\maketitle

\begin{abstract}
We consider the gravitational force exerted on a point-like perturber of mass $M$ travelling within a uniform gaseous, opaque medium at constant velocity $\mathbfit V$. The perturber irradiates the surrounding gas with luminosity $L$. The diffusion of the heat released is modelled with a uniform thermal diffusivity $\chi$. Using linear perturbation theory, we show that the force exerted by the perturbed gas on the perturber differs from the force without radiation (or standard dynamical friction). Hot, underdense gas trails the mass, which gives rise to a new force component, the heating force, with direction $+\mathbfit V$, thus opposed to the standard dynamical friction. In the limit of low Mach numbers, the heating force has expression $F_\mathrm{heat}=\gamma(\gamma-1)GML/(2\chi c_s^2)$, $c_s$ being the sound speed and $\gamma$ the ratio of specific heats. In the limit of large Mach numbers, $F_\mathrm{heat}=(\gamma-1)GML/(\chi V^2)f(r_\mathrm{min}V/4\chi)$, where $f$ is a function that diverges logarithmically as $r_\mathrm{min}$ tends to zero.  Remarkably, the force in the low Mach number limit does not depend on the velocity.  The equilibrium speed, when it exists, is set by the cancellation of the standard dynamical friction and heating force. In the low Mach number limit, it scales with the luminosity to mass ratio of the perturber.  Using the above results suggests that Mars- to Earth-sized planetary embryos heated by accretion in a gaseous protoplanetary disc should have eccentricities and inclinations that amount to a sizeable fraction of the disc's aspect ratio, for conditions thought to prevail at a few astronomical units.
\end{abstract}


\section{Introduction}
\label{sec:introduction}

A massive perturber moving across a medium triggers a disturbance in
this medium, by gravity. Dynamical friction is the back reaction of
this disturbance on the perturber itself.  It is directed against the
motion of the perturber, which it tends to slow down. Initially worked
out for collisionless media by \citet{1943ApJ....97..255C}, it was
subsequently studied in gaseous media
\citep{1980ApJ...240...20R,1999ApJ...513..252O}.  It has been
initially suggested that in the subsonic regime in gaseous media there
would be no net force exerted on the perturber
\citep{1980ApJ...240...20R}, owing to the symmetry of the perturbed
density field with respect to the perturber, in steady state. However,
\citet{1999ApJ...513..252O} showed that a time-dependent analysis is
required to correctly evaluate the net force in the subsonic regime:
the disturbance is at all times contained within a sonic sphere
centred on the initial location of the perturber, which induces
asymmetries in the far-field with respect to the actual position of
the perturber. These asymmetries ultimately yield a constant net
force, which increases with the Mach number, and scales with the
velocity of the perturber in the limit of a low Mach number. This
analysis, performed in the framework of linear perturbation theory,
was subsequently corroborated by the non-linear numerical simulations
of \citet{1999ApJ...522L..35S}.

In the supersonic regime, the force is inversely proportional to the
square of the velocity in the limit of a large Mach number. Also, in
this regime, the force scales with the logarithm of the ratio of the
maximal to minimal distance over which the medium is perturbed. The
divergence of this quantity at large scale is suppressed if the
perturber moves in a slab or protoplanetary disc
\citep{pap2002,2011ApJ...737...37M,2013ApJ...762...21C}. The minimal
distance has been found to be of the order of the gravitational radius
$GM/V^2$ by \citet{2011MNRAS.418.1238C}, who performed non-linear
calculations in the ballistic approximation.

Dynamical friction can be used to study some aspects of planet-disc
tidal interactions \citep{pap2002,2011ApJ...737...37M}. While the time
evolution of orbital elements of low-eccentricity objects must be
obtained by a resonant torque calculation, the behaviour of objects
with an eccentricity larger than the disc's aspect ratio can be
obtained through a dynamical friction calculation. In this regime
indeed, the disc's response becomes local and the Keplerian shear can
be neglected.

It has been recently shown that the modification of the disc's
structure by a hot protoplanet that irradiates its immediate
surroundings can significantly alter the tidal force from the disc,
with a sizeable impact on planetary migration
\citep{2015Natur.520...63B}. This comes from the fact that the force
arising from the disturbance due to the heat release has the same
order of magnitude as the force due to the disturbance induced by the
planet's gravity. Motivated by the serendipitous coincidence between
the order of magnitude of these two forces, we investigate here the
impact of heat release in the simpler context of dynamical friction,
using linear perturbation theory. We lay down our equations in
section~\ref{sec:governing-equations}, and work out the force arising
from heat release in the limits of low Mach numbers (section
\ref{sec:heating-force-limit}) and large Mach numbers
(section~\ref{sec:heating-force-limit-1}). We then discuss the
behaviour of a perturber subjected to the sum of the heating force and
standard dynamical friction in section~\ref{sec:total-force-pert}.  We
finally discuss the case of planetary embryos and protoplanets heated
by accretion, embedded in opaque protoplanetary discs, in section
\ref{sec:discussion}, and sum up our findings in
section~\ref{sec:conclusion}.

\section{Governing equations}
\label{sec:governing-equations}
We consider the linearised equations for the perturbations of density $\rho$,
energy density $e$ and velocity $\mathbfit v$ of an ideal gas with uniform
unperturbed density $\rho_0$, energy density $e_0$ and velocity $\mathbfit
v_0$, subjected to the gravity of a point-like mass $M$, located at the
position $r_0$, and releasing energy at the constant rate $L$:
\begin{align}
\label{eq:1}
  \partial_t\rho+\mathbfit v_0\cdot\nabla\rho+\rho_0\nabla\cdot\mathbfit v&=0\\
\label{eq:2} 
  \partial_t\mathbfit v+(\mathbfit v_0\cdot\nabla)\mathbfit v+\frac{\nabla p}{\rho_0}&=-\nabla\Phi\\
\label{eq:3} 
  \partial_t\mathbfit e+\mathbfit
  v_0\cdot\nabla e+\gamma e_0\nabla\cdot\mathbfit
  v+\frac{e_0}{\rho_0}\chi\Delta \rho-\chi\Delta e&=L\delta(\mathbfit
       r-\mathbfit r_0),
\end{align}
where
\begin{equation}
  \label{eq:4}
  \Phi=-\frac{GM}{|\mathbfit r-\mathbfit r_0|}
\end{equation}
is the gravitational potential of the mass, $\chi$ is the thermal
diffusivity of the gas, $\gamma$ is the ratio of specific heats and
$p$ is the pressure. The last two terms of the left hand side of
Eq.~\eqref{eq:3} come from the linearisation of the divergence of the
energy flux density $\mathbfit q $ arising from heat conduction and
given by Fourier's law:
\begin{equation}
  \label{eq:5}
  \mathbfit q = -c_v\rho_0\chi\nabla T,
\end{equation}
where $c_v$ is the specific heat at constant volume and where the
temperature $T$ is given by $T=(e_0+e)/[c_v(\rho_0+\rho)]$.
The perturbations of energy density and pressure are related by:
\begin{equation}
  \label{eq:6}
  p=(\gamma-1)e
\end{equation}
and a similar relationship holds for the unperturbed quantities. We
denote $c_s$ the adiabatic sound speed:
\begin{equation}
  \label{eq:7}
  c_s=\sqrt{\frac{\gamma(\gamma-1)e_0}{\rho_0}}.
\end{equation}
Assuming the perturber is introduced at $t=0$, we write
$\rho_{M,L}(\mathbfit r,t)$ the solution of the linear system of
Eqs.~\eqref{eq:1}-\eqref{eq:3}, and $\mathbfit F_{M,L}(t)$ the force
exerted by the gas on the perturber:
\begin{equation}
  \label{eq:8}
  \mathbfit F_{M,L}(t) = GM\iiint \frac{\rho_{M,L}(\mathbfit r, t)(\mathbfit r-\mathbfit
    r_0)}{|\mathbfit r-\mathbfit r_0|^3}d^3\mathbfit r.
\end{equation}
We have:
\begin{equation}
  \label{eq:9}
  \rho_{M,L}(\mathbfit r,t)=\rho_{M,0}(\mathbfit r,t)+\rho_{0,L}(\mathbfit r,t),
\end{equation}
which reads that the density perturbation $\rho_{M,L}$ due to a
massive and luminous perturber of mass $M$ and luminosity $L$ is the
sum of the density perturbations respectively due to a massive,
non-luminous perturber ($\rho_{M,0}$) and to a massless, luminous
perturber ($\rho_{0,L}$). We write:
\begin{equation}
  \label{eq:10}
  \mathbfit F_{M,L}(t) = \mathbfit F_\mathrm{DF} (t)+\mathbfit F_\mathrm{heat} (t),
\end{equation}
where $\mathbfit F_\mathrm{DF}(t)$ and $\mathbfit F_\mathrm{heat}(t)$
are the forces arising respectively from the density distributions
$\rho_{M,0}(\mathbfit r,t)$ and $\rho_{0,L}(\mathbfit r,t)$. The
expression of $\mathbfit F_\mathrm{DF}(t)$, which we call hereafter
the standard dynamical friction, has been worked out by
\citet{1999ApJ...513..252O} for a perturber in an adiabatic gas. We
seek here an expression for $\mathbfit F_\mathrm{heat}(t)$ in steady
state, which we call the heating force. In
sections~\ref{sec:heating-force-limit}
and~\ref{sec:heating-force-limit-1} below we work out the heating
force alone, respectively in the limits of low and large Mach
numbers. Subsequently we give an estimate of the total force in
section~\ref{sec:total-force-pert}.

\section{Heating force in the limit of a low Mach number}
\label{sec:heating-force-limit}
\subsection{Perturbation of density field}
We place ourselves in the perturber frame and set
$\mathbfit v_0=-\mathbfit V$, $\mathbfit r_0=\mathbfit 0$ in
Eqs.~\eqref{eq:1}-\eqref{eq:3}, where $\mathbfit V$ it the velocity of
the perturber with respect to the gas. We use a Cartesian frame
$(x,y,z)$ such that $\mathbfit V=V\mathbf{\hat z}$, and we assume a
steady state. Eqs.~\eqref{eq:1}-\eqref{eq:3} can be recast as:
\begin{align}
\label{eq:11}
  -V\partial_z\rho+\rho_0\nabla\cdot \mathbfit v&=0\\
\label{eq:12}
  -V\partial_zv_j+\frac{\partial_jp}{\rho_0}&=0\mbox{~~}j=1,2,3\\
\label{eq:13}
  -V\partial_zp+\gamma p_0\nabla\cdot\mathbfit 
  v+\frac{c_s^2}{\gamma}\chi\Delta\rho-\chi\Delta 
  p&=(\gamma-1)L\delta(\mathbfit r) 
\end{align}
We denote with a tilde the three-dimensional Fourier transform,
adopting the following conventions of sign and normalisation:
\begin{eqnarray}
  \tilde{\rho}(\mathbfit k)&=&\iiint 
  \rho(\mathbfit r)e^{-i\mathbfit k\cdot\mathbfit r}d^3\mathbfit r \nonumber\\
  \rho(\mathbfit r)&=&\frac{1}{(2\pi)^3}\iiint 
  \tilde{\rho}(\mathbfit k)e^{i\mathbfit k\cdot\mathbfit r}d^3\mathbfit k 
  \label{eq:14}
\end{eqnarray}
Eq.~\eqref{eq:12} becomes:
\begin{equation}
  \label{eq:15}
Vk_z\tilde{v}_j=k_j\frac{\tilde{p}}{\rho_0}  .
\end{equation}
Multiplying Eq.~\eqref{eq:15} by $k_j$ and summing on $j$, then using
Eq.~\eqref{eq:11}, we obtain the relation:
\begin{equation}
  \label{eq:16}
  k^2\tilde{p}=V^2k_z^2\tilde{\rho},
\end{equation}
where $k^2=k_x^2+k_y^2+k_z^2$.
Using Eqs.~\eqref{eq:16} and~\eqref{eq:11} to substitute respectively
the pressure and the divergence of velocity in Eq.~\eqref{eq:13}, we
obtain:
\begin{equation}
  \label{eq:17}
  \tilde{\rho}(k_x,k_y,k_z)=\frac{(\gamma-1)L}{\chi\left(k_z^2V^2-\frac{c_s^2}{\gamma}k^2\right)+i\left(c_s^2-\frac{k_z^2V^2}{k^2}\right)k_zV}.
\end{equation}
 In the limit $(V/c_s)^2\ll 1/\gamma$,
$\tilde\rho$ can be approximated by
\begin{equation}
  \label{eq:18}
  \tilde{\rho}(k_x,k_y,k_z)=-\frac{\gamma(\gamma-1)L/c_s^2}{\chi
    k^2-i\gamma k_zV}.
\end{equation}
We introduce the function:
\begin{equation}
  \label{eq:19}
  R(\mathbfit r)=\rho(\mathbfit r) e^{z\gamma V/2\chi},
\end{equation}
which has the following Fourier transform:
\begin{align}
\nonumber
  \tilde R(k_x,k_y,k_z)&=\tilde\rho\left(k_x,k_y,k_z+i\frac{\gamma
                         V}{2\chi}\right)\\
\label{eq:20}
&=-\frac{\gamma(\gamma-1)L/c_s^2}{\chi k^2+\frac{\gamma^2V^2}{4\chi}}.
\end{align}
The inverse Fourier transform of Eq.~\eqref{eq:20} yields
\begin{equation}
  \label{eq:21}
  R(\mathbfit r)=-\frac{\gamma(\gamma-1)Le^{-r\gamma |V|/2\chi}}{4\pi \chi c_s^2 r},
\end{equation}
and thus
\begin{equation}
  \label{eq:22}
  \rho(\mathbfit r)=-\frac{\gamma(\gamma-1)L}{4\pi \chi c_s^2 r}e^{-z \gamma V/2\chi}e^{-r\gamma |V|/2\chi}.
\end{equation}
On the $z$-axis ($r=|z|$), in the upstream flow ($z\,\mbox{sign}(V)>0$), the
density perturbation follows the law:
\begin{equation}
  \label{eq:23}
  \rho(0,0,z)=-\frac{\gamma(\gamma-1)L}{4\pi\chi c_s^2
    r}e^{-|z|/\lambda},
\end{equation}
where the cut-off distance 
\begin{equation}
  \label{eq:24}
  \lambda=\chi/\gamma|V|
\end{equation}
is set by the competition between diffusion and advection and
represents the typical distance over which the heat can diffuse
against the headwind. Remarkably, the density perturbation on the
$z$-axis in the downstream flow ($z\,\mbox{sign}(V)<0$) follows a law
that is indistinguishable from the law without headwind:
\begin{equation}
  \label{eq:25}
  \rho(0,0,z)=-\frac{\gamma(\gamma-1)L}{4\pi\chi c_s^2 r}.
\end{equation}
We will hereafter refer to $\lambda$ as the cut-off distance or, more
informally, as the size of the hot plume, since it represents the
typical size of the perturbed region.

\subsection{Force expression}
\label{sec:force-expression}
We substitute Eq.~\eqref{eq:22} in~\eqref{eq:8}, which we express
using spherical coordinates. Without loss of generality, we assume
here $V>0$. This yields:
\begin{equation}
\label{eq:26}
  F_\mathrm{heat}=-\frac{\gamma(\gamma-1)GML}{2\chi c_s^2}I,
\end{equation}
where $I$ is given by:
\begin{equation}
\label{eq:27}
I=\int_0^\infty
  \int_0^\pi\frac 1r\cos\theta
  e^{r/2\lambda(-1-\cos\theta)}\sin\theta d\theta dr
\end{equation}
We perform the change of variable $u=\cos\theta$ and integrate by
parts in $u$, which yields
\begin{equation}
\label{eq:28}
I=-\int_0^\infty\frac{1+e^{-2r'}}{r'^2}-\frac{1-e^{-2r'}}{r'^3}dr',
\end{equation}
with $r'=r/2\lambda$. We have:
\begin{equation}
\label{eq:29}
I=-\left.\frac{1-2r'-e^{-2r'}}{2r'^2}\right|_0^\infty=-1.
\end{equation}
If we now relax the assumption $V>0$, then using the symmetry
$V\rightarrow -V$, $\rho(x,y,z)\rightarrow \rho(x,y,-z)$, we finally
express the heating force, in the regime of low Mach numbers, as:
\begin{equation}
  \label{eq:30}
  F_\mathrm{heat}=\frac{\gamma(\gamma-1)GML}{2\chi c_s^2}\mbox{sign}(V).
\end{equation}
\begin{figure}
  \includegraphics[width=\columnwidth]{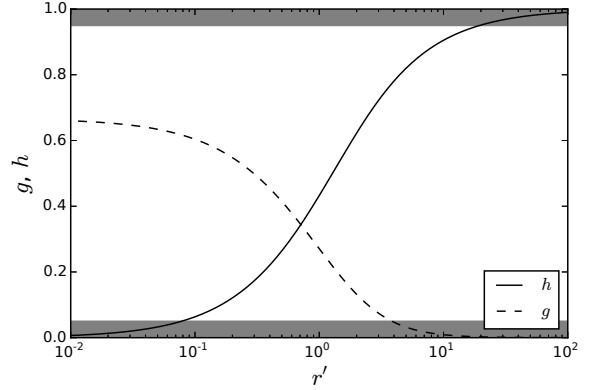}
  \caption{Graphs of functions $g$ (dashed line) and $h$ (solid line),
    respectively given by Eqs.~\eqref{eq:31} and~\eqref{eq:32}. The  
    greyed areas represent the innermost and outermost five percent  
    contributions, which are respectively obtained for $r'=0.078$ and  
    $r'=19.5$.}
\label{fig:forcesub}
\end{figure} 
It is therefore directed along the direction of motion, and has a
magnitude independent of $V$ and $\rho_0$. We
show in Fig.~\ref{fig:forcesub} the integrand of Eq.~\eqref{eq:28}:
\begin{equation}
  \label{eq:31}
  g(r')=\frac{1+e^{-2r'}}{r'^2}-\frac{1-e^{-2r'}}{r'^3}
\end{equation}
and its integral that vanishes for $r'\rightarrow 0$:
\begin{equation}
  \label{eq:32}
  h(r')=\frac{1-2r'-e^{-2r'}}{2r'^2}+1.
\end{equation}
The former shows the contribution of spherical shells to the net
force and the latter shows the cumulative contribution of all
material within radius $r$ (in units of $2\lambda$). This shows that
approximately $40$~\% of the force arises from the sphere
$r<2\lambda$.

\subsection{Condition for linearity}
\label{sec:condition-linearity}
The assumption that the density perturbation can be given by a linear
analysis breaks down when $\rho$, given by Eq.~\eqref{eq:22}, has a
value comparable to $\rho_0$. This happens for $r<r_\mathrm{NL}^\mathrm{heat}$, with
\begin{equation}
  \label{eq:33}
  r_\mathrm{NL}^\mathrm{heat}=\frac{\gamma(\gamma-1)L}{4\pi\chi c_s^2\rho_0}.
\end{equation}
As long as this radius is small compared to the cut-off distance
(which is the characteristic length scale over which the force is
produced), Eq.~\eqref{eq:26} provides a correct value of the heating
force. The condition for linearity can therefore be written as:
\begin{equation}
  \label{eq:34}
  L\ll \frac{4\pi\chi^2c_s^2\rho_0}{\gamma^2(\gamma-1)|V|}.
\end{equation}
Introducing the critical luminosity
\begin{equation}
  \label{eq:35}
  L_c=\frac{4\pi}{\gamma^2(\gamma-1)}\chi^2c_s\rho_0,
\end{equation}
this condition can be recast as:
\begin{equation}
  \label{eq:36}
  L\ll \frac{L_c}{{\cal M}},
\end{equation}
where ${\cal M}$ is the Mach number.

\subsection{Response time}
\label{sec:time-scale-appe}
The time for the heating force to develop corresponds to the time it
takes to establish the hot, underdense plume. This, in order of
magnitude, is also the time it takes for the heat to diffuse over a
length scale $2\lambda$, which is:
\begin{equation}
  \label{eq:37}
  \tau\sim\frac{(2\lambda)^2}{4\chi}=\frac{\chi}{\gamma^2V^2}.
\end{equation}
We use $2\lambda$ in this expression since this roughly corresponds to
length scale within which half of the heating force arises.
We note that $\tau\rightarrow\infty$ when $V\rightarrow 0$. In the
special case $V=0$, no heating force ever appears on the perturber,
for symmetry reasons. This is a singular case, however: as soon as the
perturber has a finite velocity, however small, the heating force
eventually develops along the direction of motion.

\section{Heating force in the limit of a large Mach number}
\label{sec:heating-force-limit-1}

In this section we place ourselves in the frame at rest with the
gas. For the sake of definiteness we assume $V>0$, and we assume that
at the instant at which we evaluate the force, the perturber is at
$z=0$. The perturber deposits an amount of energy $Ldz/V$ in each
elementary interval $[z,z+dz]$, over the whole negative $z$-axis. We
firstly evaluate the force due to the disc's response to a singular
release of energy at a given value of $z$, then we integrate over $z$
to get the force arising from the perturbation due to an energy
release over all the previous positions of the perturber. The time
elapsed between the passage of the perturber at a position $z$ and the
time at which it reaches the origin is $t=|z|/V\ll|z|/c$. We are
therefore interested, when evaluating the force at a distance $r$ and
a time $t$ after a singular energy release, in the case $r\gg ct$.

\subsection{Force arising from a singular energy deposition}
\label{sec:force-arising-from}
The perturbed density has at any time spherical symmetry about the
point of release. By virtue of Gauss's theorem, it suffices to know
the variation of the mass enclosed in the sphere centred on the point
of release and containing the point where the force is to be
evaluated. Here we assume the release to occur at the origin and at
$t=0$, and we seek the variation of mass $\Delta M(r,t)$.  This is
tantamount to solving Eqs.~\eqref{eq:1} to~\eqref{eq:3} with
$\mathbfit v_0=\mathbfit 0$, $L=E\delta(t)$ and $M=0$, $E$ being the
amount of energy released. Denoting with a bar the space and time
Fourier transform:
\begin{equation}
  \bar{e}(\mathbfit k,\omega)=\iiiint 
  e(\mathbfit r,t)e^{-i(\mathbfit k\cdot\mathbfit r-\omega t)}d^3\mathbfit r\,dt,
  \label{eq:38}
\end{equation}
Eq.~\eqref{eq:2} yields:
\begin{equation}
  \label{eq:39}
  \omega\bar{v_j}=k_j\frac{\bar{p}}{\rho_0}.
\end{equation}
Multiplying Eq.~\eqref{eq:39} by $k_j$ and summing on $j$, we obtain:
\begin{equation}
  \label{eq:40}
  \bar{\rho}=\frac{(\gamma-1)k^2}{\omega^2}\bar{e}.
\end{equation}
Using this relation, we can infer from Eqs.~\eqref{eq:1} and~\eqref{eq:3}:
\begin{equation}
  \label{eq:41}
  \bar{e}=\frac{E}{\chi k^2\left(1-\frac{k^2c_s^2}{\gamma\omega^2}\right)+i\left(\frac{c_s^2k^2}{\omega}-\omega\right)}.
\end{equation}
Since we are interested in the response at $r\gg ct$, the Fourier
components that contribute to the response fulfil $kc_s\ll\omega$,
hence:
\begin{equation}
  \label{eq:42}
  \bar{e}\approx\frac{E}{\chi k^2-i\omega}.
\end{equation}
This corresponds to the solution of the simple diffusion equation
\begin{equation}
  \label{eq:43}
\partial_te-\chi\Delta e=E\delta(t)\delta(\mathbfit r),
\end{equation}
which is:
\begin{equation}
  \label{eq:44}
  e(r,t)=\frac{E}{8\pi^{3/2}(\chi t)^{3/2}}e^{-r^2/4\chi t}.
\end{equation}
This implies, using Eq.~\eqref{eq:6} and Eq.~\eqref{eq:2} with $\mathbfit v_0=0$ and
$\Phi=0$:
\begin{equation}
  \label{eq:45}
  \partial_tv=\frac{(\gamma-1)Er}{16\rho_0\pi^{3/2}(\chi t)^{5/2}}e^{-r^2/4\chi t},
\end{equation}
and upon integration from $0$ to $t$:
\begin{equation}
  \label{eq:46}
  v(t)=\int_0^t\partial_tvdt=\frac{(\gamma-1)Er^{-2}}{2\rho_0\pi^{3/2}\chi}\Gamma\left(\frac
    32,\frac{r^2}{4\chi t}\right),
\end{equation}
where $\Gamma$ is the upper incomplete gamma function. The
mass variation in the sphere of radius $r$ centred on the release
point is obtained by another integration:
\begin{align}
  \Delta M(r,t)&=-\int_0^t4\pi r^2\rho_0v(t)dt\nonumber\\
  &=-\frac{2(\gamma-1)E}{\sqrt\pi\chi}\int_0^t\Gamma\left(\frac
    32,\frac{r^2}{4\chi t}\right)dt\nonumber\\
  \label{eq:47}
&=-\frac{(\gamma-1)Er^2}{2\sqrt\pi\chi^2}\mu\left(\frac{r}{\sqrt{4\chi t}}\right)
\end{align}
\begin{figure}
  \includegraphics[width=\columnwidth]{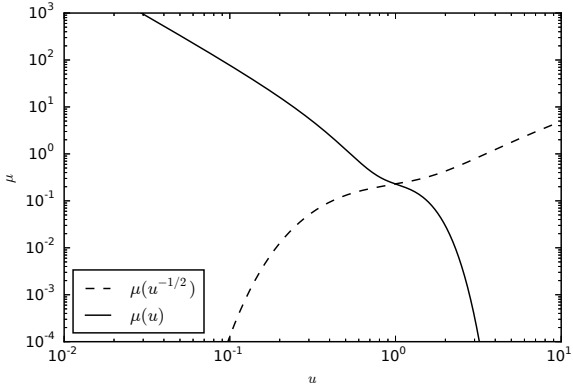}
  \caption{Graphs of the functions $u\mapsto\mu(u)$ (solid line) and
    $u\mapsto\mu(u^{-1/2})$ (dashed line).}
\label{fig:super1}
\end{figure} 
where the function $\mu$ is defined by:
\begin{equation}
  \label{eq:48}
  \mu(u)=u\exp(-u^2)+\frac{\sqrt\pi}{2}(u^{-2}-2)\,\mathrm{erfc}(u).
\end{equation}
Finally, the force arising from this disturbance has the magnitude
\begin{equation}
  \label{eq:49}
  F(r,t)=-\frac{GM\Delta M(r,t)}{r^2}=\frac{(\gamma-1)GME}{2\sqrt\pi\chi^2}\mu\left(\frac{r}{\sqrt{4\chi t}}\right)
\end{equation}
and is directed opposite the point of release. Fig.~\ref{fig:super1}
shows the graph of $\mu(u)$ and $\mu(u^{-1/2})$ which can respectively
be seen as the normalised force as a function of $r$, in units of
$r_\mathrm{diff}=\sqrt{4\chi t}$, and as the normalised force as a
function of $t$, in units of $t_\mathrm{diff}=r^2/4\chi$. A sharp
cut-off of the force is found for $t<t_\mathrm{diff}$ (or,
equivalently, for $r>r_\mathrm{diff}$). The time scale
$t_\mathrm{diff}$ is indeed the time it takes for the disturbance
created by the energy release to reach the radius~$r$.

\subsection{Net heating force}
\label{sec:net-heating-force}
We can now complete the procedure outlined at the beginning of
section~\ref{sec:heating-force-limit-1} by integrating the force
expression of Eq.~\eqref{eq:49}, substituting $E$ with $Ldz/V$ and $t$
with $r/V$. In the integral $r$ represents $-z$, where $z$ runs
through the whole set of positions previously occupied by the
perturber. We do not integrate all the way to $r=0$ however, but
rather introduce a minimal radius of integration $r_\mathrm{min}$, the
physical meaning of which will be specified later, as
the force integral is found to diverge as $r_\mathrm{min}$ tends to
zero.  Since the force arising from each elementary release is
directed in the direction opposite to the release, as shown in
section~\ref{sec:force-arising-from}, and since the perturber has
released energy over the negative $z$-axis, the net force is
positive. It has the expression:
\begin{align}
  F_\mathrm{heat}&=-\int_{r_\mathrm{min}}^\infty\frac{GM\Delta
                   M(r,r/V)}{r^2}dr\nonumber\\
&=\frac{(\gamma-1)GML}{2\sqrt{\pi}V\chi^2}\int_{r_\mathrm{min}}^\infty\mu\left(\frac{rV}{4\chi}\right)dr\nonumber\\
  \label{eq:50}
&=\frac{2(\gamma-1)GML}{\sqrt{\pi}\chi
  V^2} f\left(\frac{r_\mathrm{min}V}{4\chi}\right),
\end{align}
where $f$ is given by:
\begin{equation}
  \label{eq:51}
  f(\epsilon)=\int_\epsilon^\infty\mu(\sqrt u)du.
\end{equation}
Since $\mu(\sqrt u)\propto 1/u$ when $u$ tends to zero, $f$ diverges
logarithmically as $\epsilon$ tends to zero. After some cumbersome
manipulations, we can give an equivalent of $f(\epsilon)$ in the limit
$\epsilon\rightarrow 0$:
\begin{equation}
  \label{eq:52}
  f(\epsilon) \approx \frac{\sqrt\pi}{2}(-\gamma-2\log 2-\log\epsilon),
\end{equation}
where, in this expression only, $\gamma\approx0.577$ represents the
Euler-Mascheroni constant. Note that Eq.~\eqref{eq:52} is an
approximate expression valid only when $\epsilon\ll 1$, and that
$f(\epsilon)$ is positive for any value of $\epsilon$, since the
integrand of the right hand side of Eq.~\eqref{eq:51} is positive.
\begin{figure}
  \includegraphics[width=\columnwidth]{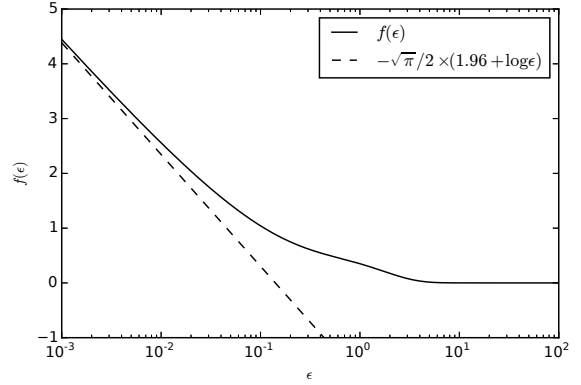}
  \caption{Graphs of the functions $f$ (solid line) and
   its approximation given by Eq.~\eqref{eq:52}.}
\label{fig:supersonicf}
\end{figure}
Fig.~\ref{fig:supersonicf} shows the graphs of the function $f$ and
its approximation. The latter gives a value reasonably close to
$f$ for $\epsilon \la 10^{-2}$.  Using Eqs.~\eqref{eq:50}
and~\eqref{eq:52}, we get the force expression in the limit
$r_\mathrm{min}\ll 4\chi/V$:
\begin{equation}
  \label{eq:53}
  F_\mathrm{heat}=\frac{(\gamma-1)GML}{\chi V^2}\left[-1.96-\log\left(\frac{r_\mathrm{min}V}{4\chi}\right)\right]
\end{equation}

\subsection{Magnitude of the minimal radius}
\label{sec:magn-minim-radi}
We do not undertake here a precise calculation of the minimal radius
$r_\mathtt{min}$ to be used in Eq.~\eqref{eq:53}. The heating force in
the supersonic regime only weakly depends on this quantity. It
corresponds to the minimal distance from the perturber at which the
asymmetry in the gas density appears. It should correspond either to
the body's physical size, or to the mean free path of photons (when
radiative transfer is the origin of heat diffusion), whichever is
larger. This last quantity is by far the largest in the application
case that we will consider in section~\ref{sec:discussion}.

\subsection{Response time}
\label{sec:time-scale-appe-1}
The force in the supersonic case arises from the diffusion of
disturbances that beat the perturber, i.e. disturbances produced at a
distance $r$ that verifies $r^2/4\chi < r/V$, or $r < r_c=4\chi/V$.
The time scale to develop the force is therefore
\begin{equation}
  \label{eq:54}
  \tau'=\frac{r_c}{V}=\frac{4\chi}{V^2}.
\end{equation}
Within a numerical coefficient, this time scale is comparable to the
time scale in the subsonic regime, given by Eq.~\eqref{eq:37}.

\subsection{Condition for linearity}
\label{sec:condition-linearity-1}
We can consider that the perturbation arises from a region of  size
$r_c$ next to the perturber, with a typical perturbation of density
$\Delta \rho$, which gives rise to the force:
\begin{equation}
  \label{eq:55}
  F\sim GM\Delta \rho r_c.
\end{equation}
Equating this expression with the expression of
Eq.~\eqref{eq:53} and dropping dimensionless factors of order
unity, we can express the condition for linearity $\Delta\rho \ll
\rho_0$ as
\begin{equation}
  \label{eq:56}
  L \ll L_c{\cal M},
\end{equation}
where $L_c$ is defined at Eq.~\eqref{eq:35}.

\section{Total force on the perturber}
\label{sec:total-force-pert}
As said in section~\ref{sec:governing-equations}, the total
force on the perturber, in the linear regime, is the sum of the
standard dynamical friction and of the heating force.
When the perturber's gravity is taken into account, the flow becomes
non-linear over the Bondi radius
\begin{equation}
  \label{eq:57}
  R_B=\frac{GM}{c_s^2}.
\end{equation}
Our linear analysis fails within the Bondi radius. The singular
release of energy considered in section~\ref{sec:governing-equations}
should be replaced by an energy release at the Bondi radius, assumed
to be small compared to the other length scales of the problem. The
heat released by the planet at the centre of the Bondi sphere can
diffuse outwards or can trigger acoustic waves which can redistribute
the mass within the Bondi sphere. Most of the heat released at the
centre by the planet will emerge of the Bondi sphere as an excess of
internal energy, as required to trigger the disc's response seen in
the previous sections, if the diffusion timescale over the Bondi
radius is smaller than the acoustic timescale over the Bondi radius:
\begin{equation}
  \label{eq:58}
  \frac{R_B^2}{4\chi}\ll\frac{R_B}{c_s},
\end{equation}
which can be recast as:
\begin{equation}
  \label{eq:59}
  M\ll\frac{4c_s\chi }{G}.
\end{equation}
We define the critical mass $M_c$ as
\begin{equation}
  \label{eq:60}
  M_c=\frac{c_s\chi}{G},
\end{equation}
and we assume in the following that the dimensionless ratio
$GM/c_s\chi =M/M_c$ is lower than one. We note that this assumption also
implies that the cut-off distance~$\lambda$ is larger than the Bondi radius.

\subsection{Regime of low Mach numbers}
\label{sec:regime-low-mach}
We specialise here to the case of low Mach numbers. In this regime,
the standard dynamical friction has the expression
\citep{1999ApJ...513..252O}:
\begin{equation}
  \label{eq:61}
  F_\mathrm{DF}=-\frac{4\pi(GM)^2\rho_0V}{3c_s^3},
\end{equation}
In the work of \cite{1999ApJ...513..252O}, thermal diffusion is not
considered and the gas is adiabatic. We have not analysed the standard
dynamical friction in the presence of thermal diffusion. It is
reasonable to assume that in the limit of a small thermal diffusion,
the gas behaves adiabatically and Eq.~\eqref{eq:61} holds, whereas
when thermal diffusion is large, it behaves isothermally, and the
isothermal sound speed should be used in stead of the adiabatic one in
Eq.~\eqref{eq:61}, yielding a force larger than the adiabatic estimate
by a factor $\gamma^{3/2}$. For the sake of definiteness we assume
hereafter that the thermal diffusivity is sufficiently small that the
adiabatic sound speed can be used in Eq.~\eqref{eq:61}. Since the
actual sound speed cannot differ from the adiabatic one by a large
factor, we expect Eq.~\eqref{eq:61} to provide a correct order of
magnitude of the standard dynamical friction in our case.

In order for the net force to be given by the sum of the linear
estimates of the dynamical friction and heating force, the relative
perturbation of the density must be small over the cut-off distance
$\lambda$ given by Eq.~\eqref{eq:24}. This, along with the condition
expressed by Eq.~\eqref{eq:34}, requires that the Bondi radius is much
smaller than $\lambda$, a condition that is automatically satisfied
when the inequality of Eq.~\eqref{eq:59} is verified, since $V< c$.

\subsection{Equilibrium speed}
\label{sec:equilibrium-speed}
When the two conditions $r_\mathrm{NL}^\mathrm{heat} \ll \lambda$ and
$R_B \ll \lambda$ are fulfilled, the net force
is given by:
\begin{equation}
  \label{eq:62}
  F=F_\mathrm{heat}+F_\mathrm{DF}=\frac{\gamma(\gamma-1)GML}{2\chi
    c_s^2}-\frac{4\pi(GM)^2\rho_0V}{3c_s^3}.
\end{equation}
It vanishes at the equilibrium speed
\begin{equation}
  \label{eq:63}
  V_0=\frac{3\gamma(\gamma-1)Lc_s}{8\pi\rho_0 GM\chi}.
\end{equation}
This speed corresponds to a steady configuration: if the perturber is
slower, it is sped up by a positive net force, and slowed down
otherwise. It is noteworthy that this equilibrium speed scales with
the sound speed and the luminosity to mass ratio of the perturber, and
increases when the gas density or thermal diffusivity decreases. We
note that the equilibrium speed is much smaller than the sound speed,
as required by our hypothesis of a low Mach number, if $L\ll L_c$,
where the critical luminosity $L_c$ is defined at Eq.~\eqref{eq:35}.

\subsection{Regime of high Mach numbers}
\label{sec:regime-high-mach}
In an infinite homogeneous medium, the dynamical friction increases
logarithmically with time, and will eventually supersede the heating
force, which comes from material in the vicinity of the perturber and
does not exhibit a logarithmic divergence at large scales. In a bound
medium such as a slab or protoplanetary disc
\citep{pap2002,2011ApJ...737...37M,2013ApJ...762...21C}, there is no
such divergence and both the dynamical friction and heating force have
similar forms, with a dimensionless factor of order unity that scales
with $|\log r_\textrm{min}|$. Since they have same dependency on the
velocity in $V^{-2}$, there cannot be an equilibrium speed as in the
subsonic case: either the heating force is larger than the dynamical
friction at all speeds, and the perturber is indefinitely accelerated,
or the heating force is smaller than the dynamical friction, and the
perturber is slowed down, albeit on a longer time than if dynamical
friction was acting alone.  This bifurcation occurs for a luminosity
set by the equality of both forces (in magnitude), which yields,
assuming that the logarithmic factors of the heating force and of the
dynamical friction are equal:
\begin{equation}
  \label{eq:64}
  L\approx \frac{4\pi}{\gamma-1}GM\chi\rho_0,
\end{equation}
or, in terms of the critical luminosity introduced at Eq.~\eqref{eq:35}
\begin{equation}
  \label{eq:65}
  L=\gamma^2L_c\left(\frac{GM}{c_s\chi}\right).
\end{equation}
Therefore an object that fulfils the condition of Eq.~\eqref{eq:59} by
a sufficiently large amount, or that has a sufficiently large initial
Mach number, can fall in the accelerated regime.

Again, we stress that standard dynamical friction has not been
investigated in a gas with finite thermal diffusivity. In the
supersonic regime, gas can be accreted over a region of typical size
$R_a=2GM/V^2$ \citep{2011MNRAS.418.1238C}. The corresponding mass flow
is $\dot M_\mathrm{gas} = \pi R_a^2V\rho_0$. The energy release due to
gas accretion has order of magnitude
$L_\mathrm{gas}\sim GM\dot M_\mathrm{gas}/R_a$.  When the disc has a
finite thermal diffusivity, the heat thus released diffuses and gives
rise to a heating force. One can check that when the perturber is
sub-critical, \emph{i.e.}  when Eq.~\eqref{eq:59} is satisfied, one
has $R_a\ll \lambda$, so that the heat release can be considered as
isotropic. One can further check that the condition for linearity of
Eq.~\eqref{eq:56} on the perturbation arising from $L_\emph{gas}$ is
satisfied and that the heating force arising from $L_\emph{gas}$ is
smaller than the standard dynamical friction. A luminosity source
intrinsic to the perturber is therefore required for the heating force
to overcome the standard dynamical friction.

\subsection{Yield}
\label{sec:yield}
The system bears some similarities with a heat engine, as it converts
part of the heat released by the perturber into work against the
dynamical friction. We calculate here what fraction of the heat
released goes into the work of the heating force. In the limit in
which this force is much larger than the dynamical friction, it is
also the yield of the conversion of the heat into macroscopic kinetic
energy of the perturber.  This yield reads
\begin{equation}
  \label{eq:66}
  \eta=\frac{VF_\mathrm{heat}}{L},
\end{equation}
which, in the regime of low Mach numbers, can be recast as:
\begin{equation}
  \label{eq:67}
  \eta=\frac{\gamma(\gamma-1)}{2}\left(\frac{GM}{c_s\chi }\right){\cal M},
\end{equation}
whereas it reads, in the limit of large Mach numbers:
\begin{equation}
  \label{eq:68}
  \eta=(\gamma-1)\left(\frac{GM}{c_s\chi}\right)f{\cal M}^{-1},
\end{equation}
where $f$ is the logarithmic factor of order unity.  These expressions
are in line with our constraint on the mass exposed at the beginning
of section~\ref{sec:total-force-pert}.  They give values lower than
one when the dimensionless factor $GM/c_s\chi$ is small.

\section{Discussion}
\label{sec:discussion}

\subsection{Disc models and assumptions}
\label{sec:eccentr-incl-plan}
The purpose of this section is to assess whether the effect we report
here could be relevant to planetary embryos or protoplanets embedded
in an opaque, gaseous protoplanetary disc.  Our analysis cannot be
applied directly to the evaluation of the heating torque
\citep{2015Natur.520...63B}, which is a torque component that appears
on planetary embryos on circular orbits when they release energy in
the surrounding protoplanetary disc. The Keplerian shear of the gas is
an essential ingredient of this effect, and it is neglected in the
present analysis. In the same spirit, the formula of dynamical
friction in a gas at rest \citep{1999ApJ...513..252O} cannot be used
to evaluate the tidal torque exerted by the disc on a planet in
circular orbit. However, it provides a useful approximation to infer
the temporal behaviour of eccentricity and inclination
\citep{pap2002,2011ApJ...737...37M, 2012MNRAS.422.3611R}, especially
when these orbital elements are large compared to the aspect ratio.

In this section we use the expression supersonic (subsonic) to refer
to embryos that have an eccentricity or inclination larger (smaller)
than the aspect ratio of the disc. We assume that the heating force
exerted by the disc on an embryo is reasonably approximated by
Eq.~\eqref{eq:53} in the supersonic case, whereas it may not
necessarily be given by Eq.~\eqref{eq:30} in the subsonic case,
because the Keplerian shear, here neglected, may be important in this
case. We will work out an estimate of the eccentricity above which the
shear should be unimportant in section~\ref{sec:rand-veloc-embry}.

The parameter space under consideration has a considerable number of
dimensions. For the sake of definiteness, we consider hereafter the
fiducial disc model of \citet{2015A&A...575A..28B}, corresponding to
their Fig.~5, for the accretion rates
$\dot M=3.5\cdot 10^{-8}$~M$_\odot$.yr$^{-1}$ and
$\dot M=8.75\cdot 10^{-9}$~M$_\odot$.yr$^{-1}$, which correspond
respectively to the ages $t=300$~kyrs and $t=1$~Myr. In these disc
models we take the conditions at $r=$~3~AU, which is the location of
the snow line at $t=300$~kyrs. There, there is a large drop in the
temperature of the disc between these two dates, which has, as we will
see, a strong impact on the different key quantities.

\begin{table}
  \caption{Parameters from the disc model of Bitsch et al. (2015) 
    at $r=3$~AU that are used in the
    numerical estimates of section~\ref{sec:discussion}. The rows
    represent respectively the midplane temperature, the surface
    density of the gas, the thermal diffusivity, the adiabatic sound
    speed, the midplane density, the aspect ratio of the disc, and the
  optical depth to the midplane
  $\tau_\mathrm{eff}=\kappa(\rho_0,T)\Sigma/2$.}
 \label{tab:disc1}
 \begin{tabular}{lcc}
  \hline
  Parameter & Value at $t=300$~kyrs & Value at $t=1$~Myr\\
  \hline 
   $T$ (K)& $195$ & $70$\\
   $\Sigma$ (g.cm$^{-2}$) & $260$ & $190$\\
   $\chi$ (cm$^2$.s$^{-1}$) & $1.7\cdot 10^{16}$ & $9.8\cdot
                                                   10^{14}$\\
   $c_s$ (cm.s$^{-1}$) & $10^5$ & $6\cdot 10^4$\\ 
   $\rho_0$ (g.cm$^{-3}$) & $4.7\cdot 10^{-11}$ & $5.8\cdot
                                                  10^{-11}$\\ 
   $h$ & 0.049 & 0.029\\
   $\tau_\mathrm{eff}$ & 120 & 47 \\
  \hline
 \end{tabular}
\end{table}

We consider planetary embryos or protoplanets with mass $M$, radius
$R$, density $\rho_p=3$~g.cm$^{-3}$, and a mass doubling time
$\tau_d=M/\dot M$. Their luminosity is therefore:
\begin{equation}
  \label{eq:69}
  L=\frac{GM^2}{R\tau_d}.
\end{equation}
When $\tau_d$ and $\rho_p$ are fixed, there is a one-to-one relationship
between the mass and luminosity. This allows us to translate critical
luminosities into critical masses, which is likely more intuitive to
most readers.  For the thermal diffusivity, we use Eq.~(16) of
\cite{2014A&A...564A.135B}, and use half the value given by the
prescription of \citet{1994ApJ...427..987B} to evaluate the opacity
$\kappa$ that is required in this equation (the fiducial model of
\citeauthor{2015A&A...575A..28B} has half the metallicity used by
\citeauthor{1994ApJ...427..987B}). We use a ratio of
specific heats $\gamma=1.4$. Finally, we use the density in the
midplane of the disc, which we denote $\rho_0$.  We list our main
parameters in Tab.~\ref{tab:disc1}.

\subsection{Condition for linearity}
\label{sec:conditions-linearity}
We write hereafter a conservative condition for linearity on the
luminosity. In the subsonic regime, we must have $L\ll L_c/{\cal M}$,
whereas we must have $L\ll L_c{\cal M}$ in the supersonic regime. We
disregard the dependency on the Mach number, which we assume to be
unity, so that we have a conservative condition on the
luminosity. Translating the latter into a critical mass through the
use of Eq.~\eqref{eq:69}, we obtain:
\begin{equation}
  \label{eq:70}
  M<\left(\frac{3}{4\pi\rho_p}\right)^\frac15\left[\frac{4\pi}{\gamma^2(\gamma-1)}\cdot\frac{\chi^2c_s\rho_0\tau_d}{G}\right]^{3/5}.
\end{equation}
For  our parameters, this gives:
\begin{eqnarray}
  M &<& 6.4\left(\frac{\tau_d}{10^5\mbox{~yrs}}\right)^{3/5}\;M_\oplus\mbox{~~~at~}t=300\mbox{~kyrs}\nonumber\\
  M &<& 0.18\left(\frac{\tau_d}{10^5\mbox{~yrs}}\right)^{3/5}\;M_\oplus\mbox{~~~at~}t=1\mbox{~Myr}\nonumber  
\end{eqnarray}     
These conditions indicate that protoplanets up to several Earth masses
trigger a linear response in the disc at early time, whereas embryos
slightly above Mars's size trigger a non-linear response at later times, once
the disc has substantially cooled.

\subsection{Critical mass}
\label{sec:critical-mass}
The critical mass given by Eq.~\eqref{eq:60} has, for our set of
parameters, the following value:
\begin{eqnarray}
  M_c &=& 4.2\;M_\oplus\mbox{~~~at~}t=300\mbox{~kyrs}\nonumber\\
  M_c &=& 0.15\;M_\oplus\mbox{~~~at~}t=1\mbox{~Myr}\nonumber  
\end{eqnarray}
Its dramatic drop at larger time is essentially due to the drop in the
disc's thermal diffusivity. Perturbers with a mass below the critical
mass are essentially subjected to the heating force worked out in
sections~\ref{sec:heating-force-limit}
and~\ref{sec:heating-force-limit-1} (in a medium without shear)
whereas they are necessarily subjected to a weaker force when their
mass is larger than the critical mass, as shown by the considerations
on the yield of section~\ref{sec:yield}. The actual magnitude of the
heating force on a perturber with a mass larger than the critical mass
is nonetheless unknown at this stage. It is noteworthy that these
critical masses roughly coincide with the upper values for a linear
response obtained in section~\ref{sec:conditions-linearity}.

\subsection{Response time}
\label{sec:time-appearance}
The velocity of an eccentric and/or inclined embryo with respect to
the surrounding gas changes orientation over a dynamical time
scale. This time scale must be substantially longer than the response
time of the heating force, given by Eq.~\eqref{eq:37} or
Eq.~\eqref{eq:54}.  In the supersonic regime, we can obtain a
conservative estimate of the response time (\emph{i.e.}, an upper
value) by using the sound speed instead of the velocity in this
equation. Conversely, the actual response time is larger than this
estimate in the subsonic regime.  Using this estimate, the condition
for a short response time is:
\begin{equation}
  \label{eq:71}
  \frac{\chi}{\gamma^2c_s^2}\la\frac 1\Omega,
\end{equation}
where $\Omega$ is the orbital frequency. Eq.~\eqref{eq:71} can be
written, dropping dimensionless factors of order unity:
\begin{equation}
  \label{eq:72}
  \lambda \la H,
\end{equation}
so the condition for a short response time is equivalent to requiring a
plume size smaller than the scale height $H$ of the disc. This, in turn,
can also be expressed as
\begin{equation}
  \label{eq:73}
  \frac{GM_c}{c_s^2}\la H,
\end{equation}
which means that the embryos with a mass equal to the critical mass,
defined at Eq.~\eqref{eq:60}, must have a Bondi radius much smaller
than the disc's scale height. This condition is satisfied by a very
large factor for the two cases considered in
section~\ref{sec:critical-mass}, namely a factor of~$10$ in the first
case, and a factor of~$50$ in the second one. This also suggests that
even for embryos with an eccentricity as small as
$h/10 \sim 4\cdot 10^{-3}$, the response time of the heating force is
shorter than the dynamical timescale, so that it can have an impact on
the embryo's eccentricity.

An embryo travelling at a sizeable fraction of the sound speed with
respect to the disc material is subjected to asymmetric heating, as
more solid material impinges on its frontward side. One can consider
that the energy is released isotropically if the rotation period of
the embryo is much shorter than the response time of the plume
(strictly speaking, this is true if the rotation axis is perpendicular
to the direction of motion; in the general case there is a residual
anisotropy that depends on the angle between the rotation axis and the
direction of motion). For our first data set (at $t=300$~kyrs) a
conservative estimate of the response time is $\tau\sim 8$~days,
whereas it is only $\tau\sim 1.6$~days for our second data set (at
$t=1$~Myr). These estimates are conservative because they are
evaluated using the sound speed in Eq.~\eqref{eq:37}. If the embryo
has a lower speed, the plume is larger and the response time
longer. These typical time scales should be compared to a rotation
period of $O(10)$~hrs. The first estimate is widely larger, but the
second one might be comparable to the rotational period of the slowest
embryos. In this last, rare case the heating force is no longer directed
along the direction of motion, but changes orientation as the embryo
rotates.

\subsection{Constraints on the cut-off distance}
\label{sec:plume-size-versus}
The heating force, in the subsonic regime, can have the expression
given by Eq.~\eqref{eq:30} only if the embryo's physical radius is
much smaller than the size of the heated region (the cut-off distance
$\lambda$). For our two data sets, this distance typically amounts to
$10^{-2}$ to $10^{-3}$~AU, or $10^5$ to $10^6$~km (these values are
again conservative estimates obtained using the sound speed in the
expression of the cut-off distance).  The plume is therefore much
larger than any embedded protoplanet. Another, more demanding
condition on the plume size is that it should be larger than the mean
free path of photons $(\rho_0\kappa)^{-1}\sim H/\tau_\mathrm{eff}$, so
that the evolution of the heat within the plume is governed by
diffusion, as assumed in our analysis. This condition is satisfied by
a factor of order~$5$ for our first data set ($t=300$~kyrs), whereas
the mean free path of photons turns out to be $\sim 3$~times larger
than the conservative (minimal) plume size for our second data set
($t=1$~Myr). Our expression of the heating force would therefore be
valid, in that second case, for embryos with Mach numbers smaller than
$1/3$.

\subsection{Critical luminosity in the supersonic regime}
\label{sec:crit-lumin-supers}
We have seen in section~\ref{sec:regime-high-mach} that a perturber
with a luminosity above the critical value given by Eq.~\eqref{eq:64}
has a heating force larger than the dynamical friction at all speeds,
and can be indefinitely accelerated. Above this luminosity threshold,
a planetary embryo with a large eccentricity should therefore become
more eccentric, rather than being circularised, since the heating force
has opposite direction to the dynamical friction. Translating
the luminosity threshold, through the use of Eq.~\eqref{eq:69}, into a
mass threshold, we are led to:
\begin{equation}
  \label{eq:74}
  M>4\pi\left(\frac{3}{\rho_p}\right)^\frac12\left(\frac{\chi\rho_0\tau_d}{\gamma-1}\right)^\frac32,
\end{equation}
which reads, for the parameters of our first data set:
\begin{equation*}
  M > 33\left(\frac{\tau_d}{10^5\mbox{~yrs}}\right)^{3/2}\;M_\oplus\mbox{~~~at~}t=300\mbox{~kyrs}
\end{equation*}    
The mass doubling time must be as short as $25$~kyrs to get a mass
threshold equal to the critical mass worked out in
section~\ref{sec:critical-mass}.  Embryos of critical mass undergoing
such a vigorous accretion, if they exist, should therefore fall in the regime of
indefinite acceleration.

We do not consider our second data set, since we have seen in
section~\ref{sec:plume-size-versus} that the plume size is smaller
than the mean free path of photons in this case, when the Mach number
exceeds $1/3$.  

The ultimate value of eccentricity or inclination reached in the
supersonic regime is beyond the scope of this work. Additional
complications arise: (i) embryos that have larger velocities with
respect to the disc should have a larger accretion rate and have their
luminosity increased; (ii) when the inclination becomes of the order
of the disc's aspect ratio, embryos only spend a fraction of their
orbit within the disc; (iii) aerodynamic drag is non-negligible for
supersonic embryos up to a few Earth masses
\citep{2012MNRAS.422.3611R}; (iv) when the Mach number increases, the
plume size decreases and can be comparable to the photons mean free
path, in which case our expression for the
heating force ceases to be valid.

\subsection{Random velocities of embryos}
\label{sec:rand-veloc-embry}
We evaluate here the equilibrium speed of embryos as given by
Eq.~\eqref{eq:63}, and regard it as the magnitude of the random
velocity with respect to their circular motion. We emphasise that this
approach is not self-consistent: the equilibrium velocity exists only
in the subsonic regime, for which the Keplerian shear is important,
whereas it has been neglected in our analysis.  Nevertheless, the Mach
number that we obtain should give an indication of whether we can
expect sizeable eccentricities or inclinations as an outcome of the
heating force. We can evaluate the eccentricity above which the shear
should be unimportant on a plume. An embryo with eccentricity $e$,
orbital radius $r$ and orbital frequency $\Omega$ has relative
velocity with respect to the gas $er\Omega$, hence a cut-off distance
$\lambda$ of order $\chi/\gamma er\Omega$. By requesting that this
cut-off distance be smaller than its distance to corotation $er$, we
obtain:
\begin{equation}
  \label{eq:75}
  e>e_c=\sqrt\frac{\chi}{\gamma\Omega r^2}.
\end{equation}
The critical value for the eccentricity in our first data set
($t=300$~kyrs) is $e_c=10^{-2}$, whereas it is $e_c=3\cdot10^{-3}$ in
our second data set ($t=1$~Myr). These values are significantly
smaller than the aspect ratio, because the length scale of the
response (\emph{i.e.} the plume size) is much smaller than the
pressure scale length. This justifies the use of the heating force
expression to assess eccentricity driving even in the subsonic
regime. Using Eqs.~\eqref{eq:63} and~\eqref{eq:69}, we obtain
\begin{equation}
  \label{eq:76}
  \frac{V_0}{c_s}=\frac{3^{2/3}\gamma(\gamma-1)M^{2/3}(4\pi\rho_p)^{1/3}}{8\pi\rho_0\chi\tau_d}.
\end{equation}
Numerically, this gives
\begin{eqnarray}
  \frac{V_0}{c_s} &\sim&
                    0.2\left(\frac{\tau_d}{10^5\mbox{~yrs}}\right)^{-1}\left(\frac{M}{1\;M_\oplus}\right)^{2/3}\mbox{~~~at~}t=300\mbox{~kyrs}\nonumber\\
  \frac{V_0}{c_s} &\sim& 2.9\left(\frac{\tau_d}{10^5\mbox{~yrs}}\right)^{-1}\left(\frac{M}{1\;M_\oplus}\right)^{2/3}\mbox{~~~at~}t=1\mbox{~Myr}\nonumber  
\end{eqnarray}    
These relations show that an Earth-sized planet could reach an
eccentricity or inclination that represents a fair fraction of the
disc's aspect ratio at $t=300$~kyrs. At $t=1$~Myr, a one Earth-mass
embryo is super-critical (see section~\ref{sec:critical-mass}), so the
relation above cannot be used for this mass. When used with the
critical mass $M=0.15$~$M_\oplus$, it yields a Mach number of
order~$0.8$, larger than the threshold that we worked out in
section~\ref{sec:plume-size-versus}. This nevertheless suggests that
critical mass embryos (typically of Mars size) should experience a
significant eccentricity or inclination excitation.

We finally compare the time scale of eccentricity driving by the
heating force to the time scale of eccentricity evolution due to
accretion of solids. The eccentricity of an embryo is significantly
modified by accretion when the embryo has accreted a solid mass
comparable to its own mass, which occurs on the mass doubling time
$\tau_d$ of the embryo. An order of magnitude of the time scale
$\tau_e$ of eccentricity excitation by the heating force can be
obtained by equating the work of the heating force in the rotating
frame to the energy in the epicyclic motion (the former quantity
corresponds to the variation of the embryo's Hamiltonian in the frame
rotating with the guiding centre's angular velocity, assumed to be
constant, while the latter is the eccentricity dependent term of this
Hamiltonian):
\begin{equation}
  \label{eq:77}
  F_\emph{heat}\cdot e\Omega r\cdot \tau_e\sim\frac12M\Omega^2(er)^2.
\end{equation}
Using Eqs.~\eqref{eq:24}, \eqref{eq:30}, \eqref{eq:57}
and~\eqref{eq:69}, assuming $e\sim h$ and
neglecting dimensionless factors of order unity, we obtain:
\begin{equation}
  \label{eq:78}
  \frac{\tau_e}{\tau_d}\sim\frac{\lambda R_p}{R_B^2}.
\end{equation}
For an Earth-mass object, the first data set of Tab.~\ref{tab:disc1}
yields a value of $0.06$ for the ratio of Eq.~\eqref{eq:78}. This
small value could be expected on the following grounds: the
equilibrium eccentricity results from a competition between the
heating force and the disc's tide. The time scale for excitation
should therefore have same order of magnitude as the time scale for
eccentricity damping by the disc's tide, known to be of order of a few
thousand years for an Earth-mass object within a disc with parameters
such as those quoted in Tab.~\ref{tab:disc1} \citep{arty93b}, whereas
the mass doubling time is here $10^5$~yrs. Similar arguments apply to
the inclination excitation. The main consequence of the accretion of
solids is thus by far the effect of the heating force, rather than the
effect of the momentum of accreted solids.

\section{Conclusion}
\label{sec:conclusion}
We show that a hot body moving in an opaque, initially uniform gas is
subjected to a gravitational force from the gas that differs from the
standard estimate of dynamical friction, which neglects the deposition
of energy in the gas by the perturber. The difference, which we call
the heating force, is directed along the direction of motion, and thus
opposed to the standard dynamical friction. We give the asymptotic
value of the heating force in the low and high Mach number regimes,
using a linear analysis. We find it to be independent of the body's
velocity in the limit of a low Mach number. As a consequence, a
motionless configuration is unstable, and the perturber is eventually
pushed by the heating force along its original direction of motion,
regardless of how small is its initial velocity.

The asymptotic or equilibrium velocity, when it exists, is reached
when the heating force cancels the standard dynamical friction. When
it does not, the heating force is larger than the standard dynamical
friction at all speeds, and the perturber is indefinitely
accelerated. In the limit of low Mach numbers, the equilibrium velocity
scales with the luminosity to mass ratio of the perturber and with the
sound speed, and is inversely proportional to the thermal diffusivity
and density of the gas. From a qualitative standpoint, one could say
that the perturber ``surfs'' on the hot trail that it creates in the
gas.
 
Although accurate derivations of the mean eccentricity and inclination
of embryos or protoplanets heated by planetesimal accretion are beyond
the scope of this work, we find that the heating force should have a
significant impact on the random velocities of planetary embryos
embedded in opaque, gaseous discs, and with mass doubling time
$O(10^5)$~yrs. The mass range of applicability of our linear analysis
depends sensitively on the disc parameters, but it is broadly in the
Earth-sized range, and shifts toward lower masses as the disc cools
and its thermal diffusivity decays. Clearly, the numerical estimates
we present here stand only as a proof of concept, and much broader
explorations of disc models and distances to the central star are
necessary to evaluate the impact of the heating force in scenarios of
planetary formation.  This effect could potentially have an impact on
the growth of oligarchs within a still optically thick gaseous
disc. Standard dynamical friction has been found to play a critical
role on the outcome of the giant impact phase of terrestrial planet
formation. Eccentricity and inclination driving by the heating force
may alter outcomes of models of super-Earths formation
\citep{2016ApJ...822...54D}.

A tentative explanation for the existence of chondrules is the passage
of solid material through the bow shocks of eccentric planetary
embryos in the solar nebula \citep{2012ApJ...752...27M}. Although the
eccentricities required to reproduce the thermal history of chondrules
are in excess of those estimated in section~\ref{sec:discussion}, the
heating force implies more dynamically excited embryos in the presence
of gas, which favours scattering, and it lengthens the damping
time scale of the large eccentricities attained through scattering
events.

The remaining eccentricities and inclinations after the gas
disappearance depend on the time behaviour of the gas density and
temperature as the gaseous disc vanishes.  The thermal diffusivity of
the disc is proportional to the cube of its temperature and inversely
proportional to the square of its midplane density. In the last stages
of the disc, when its midplane density decreases substantially but it
is still large enough to be optically thick, the thermal diffusivity
should increase again, shifting the domain of validity of our linear
analysis back toward larger, Earth-sized objects.  This could excite
the eccentricity of pairs of planets brought into mean motion
resonance by convergent migration, if they are still subjected to
planetesimal or pebble accretion (hence hot) at this stage. In the
same spirit, this could provide the initial disc of embryos and
planetesimals out of which terrestrial planets are formed with an
initial angular momentum deficit \citep[AMD,][]{2000PhRvL..84.3240L},
but further work is required to assess the values of the residual
eccentricities and inclinations and the value of the initial AMD.

We have identified a luminosity threshold above which the heating
force on a supersonic embryo should always supersede the standard
dynamical friction, independently of its velocity, leading in
principle to a considerable effect on eccentricity and inclination.  Our
numerical estimates suggest that Earth-sized bodies or below
undergoing very efficient pebble accretion \citep{2015A&A...582A.112B}
might fall into this regime.

An interesting further development would be the study of the heating
force when the perturber's velocity varies with time, as has been done
for the standard dynamical friction by
\citet{2010MNRAS.401..319N}. This would allow to quantify accurately
the response time of the heating force.

Our analysis features the dimensionless ratio $GM/c_s\chi$.
Considerations on the yield of the conversion of the energy released
by the perturber into macroscopic kinetic energy suggest that the
heating force should be weaker than given by our linear analysis when
this ratio is larger than one.  This happens when the acoustic time
across the Bondi radius of the perturber becomes smaller than the
diffusion time. The simulations of \cite{2015Natur.520...63B} show
that the heating torque is indeed cut off near 3~$M_\oplus$, which
corresponds precisely to the planetary mass for which these two
timescales coincide.

\section*{Acknowledgements}

The authors acknowledge support from UNAM's PAPIIT grant IN101616, as
well as constructive comments from Gloria Koenigsberger, Jeffrey Fung,
Pablo Ben\'\i tez-Llambay, Bertram Bitsch, Wladimir Lyra and an
anonymous referee.




\bibliographystyle{mnras}


\begin{thebibliography}{}
\makeatletter
\relax
\def\mn@urlcharsother{\let\do\@makeother \do\$\do\&\do\#\do\^\do\_\do\%\do\~}
\def\mn@doi{\begingroup\mn@urlcharsother \@ifnextchar [ {\mn@doi@}
  {\mn@doi@[]}}
\def\mn@doi@[#1]#2{\def\@tempa{#1}\ifx\@tempa\@empty \href
  {http://dx.doi.org/#2} {doi:#2}\else \href {http://dx.doi.org/#2} {#1}\fi
  \endgroup}
\def\mn@eprint#1#2{\mn@eprint@#1:#2::\@nil}
\def\mn@eprint@arXiv#1{\href {http://arxiv.org/abs/#1} {{\tt arXiv:#1}}}
\def\mn@eprint@dblp#1{\href {http://dblp.uni-trier.de/rec/bibtex/#1.xml}
  {dblp:#1}}
\def\mn@eprint@#1:#2:#3:#4\@nil{\def\@tempa {#1}\def\@tempb {#2}\def\@tempc
  {#3}\ifx \@tempc \@empty \let \@tempc \@tempb \let \@tempb \@tempa \fi \ifx
  \@tempb \@empty \def\@tempb {arXiv}\fi \@ifundefined
  {mn@eprint@\@tempb}{\@tempb:\@tempc}{\expandafter \expandafter \csname
  mn@eprint@\@tempb\endcsname \expandafter{\@tempc}}}

\bibitem[\protect\citeauthoryear{{Artymowicz}}{{Artymowicz}}{1993}]{arty93b}
{Artymowicz} P.,  1993, \mn@doi [\apj] {10.1086/173470}, \href
  {http://adsabs.harvard.edu/cgi-bin/nph-bib_query?bibcode=1993ApJ...419..166A&db_key=AST}
  {419, 166}

\bibitem[\protect\citeauthoryear{{Bell} \& {Lin}}{{Bell} \&
  {Lin}}{1994}]{1994ApJ...427..987B}
{Bell} K.~R.,  {Lin} D.~N.~C.,  1994, \mn@doi [\apj] {10.1086/174206}, \href
  {http://adsabs.harvard.edu/abs/1994ApJ...427..987B} {427, 987}

\bibitem[\protect\citeauthoryear{{Benitez-Llambay}, {Masset}, {Koenigsberger}
  \& {Szulagyi}}{{Benitez-Llambay} et~al.}{2015}]{2015Natur.520...63B}
{Benitez-Llambay} P.,  {Masset} F.,  {Koenigsberger} G.,   {Szulagyi} J.,
  2015, \mn@doi [\nat] {10.1038/nature14277}, \href
  {http://adsabs.harvard.edu/abs/2015Natur.520...63B} {520, 63}

\bibitem[\protect\citeauthoryear{{Bitsch}, {Morbidelli}, {Lega}  \&
  {Crida}}{{Bitsch} et~al.}{2014}]{2014A&A...564A.135B}
{Bitsch} B.,  {Morbidelli} A.,  {Lega} E.,   {Crida} A.,  2014, \mn@doi [\aap]
  {10.1051/0004-6361/201323007}, \href
  {http://adsabs.harvard.edu/abs/2014A%26A...564A.135B} {564, A135}

\bibitem[\protect\citeauthoryear{{Bitsch}, {Johansen}, {Lambrechts}  \&
  {Morbidelli}}{{Bitsch} et~al.}{2015a}]{2015A&A...575A..28B}
{Bitsch} B.,  {Johansen} A.,  {Lambrechts} M.,   {Morbidelli} A.,  2015a,
  \mn@doi [\aap] {10.1051/0004-6361/201424964}, \href
  {http://adsabs.harvard.edu/abs/2015A%26A...575A..28B} {575, A28}

\bibitem[\protect\citeauthoryear{{Bitsch}, {Lambrechts}  \&
  {Johansen}}{{Bitsch} et~al.}{2015b}]{2015A&A...582A.112B}
{Bitsch} B.,  {Lambrechts} M.,   {Johansen} A.,  2015b, \mn@doi [\aap]
  {10.1051/0004-6361/201526463}, \href
  {http://adsabs.harvard.edu/abs/2015A%26A...582A.112B} {582, A112}

\bibitem[\protect\citeauthoryear{{Cant{\'o}}, {Raga}, {Esquivel}  \&
  {S{\'a}nchez-Salcedo}}{{Cant{\'o}} et~al.}{2011}]{2011MNRAS.418.1238C}
{Cant{\'o}} J.,  {Raga} A.~C.,  {Esquivel} A.,   {S{\'a}nchez-Salcedo} F.~J.,
  2011, \mn@doi [\mnras] {10.1111/j.1365-2966.2011.19574.x}, \href
  {http://adsabs.harvard.edu/abs/2011MNRAS.418.1238C} {418, 1238}

\bibitem[\protect\citeauthoryear{{Cant{\'o}}, {Esquivel}, {S{\'a}nchez-Salcedo}
   \& {Raga}}{{Cant{\'o}} et~al.}{2013}]{2013ApJ...762...21C}
{Cant{\'o}} J.,  {Esquivel} A.,  {S{\'a}nchez-Salcedo} F.~J.,   {Raga} A.~C.,
  2013, \mn@doi [\apj] {10.1088/0004-637X/762/1/21}, \href
  {http://adsabs.harvard.edu/abs/2013ApJ...762...21C} {762, 21}

\bibitem[\protect\citeauthoryear{{Chandrasekhar}}{{Chandrasekhar}}{1943}]{1943ApJ....97..255C}
{Chandrasekhar} S.,  1943, \mn@doi [\apj] {10.1086/144517}, \href
  {http://adsabs.harvard.edu/abs/1943ApJ....97..255C} {97, 255}

\bibitem[\protect\citeauthoryear{{Dawson}, {Lee}  \& {Chiang}}{{Dawson}
  et~al.}{2016}]{2016ApJ...822...54D}
{Dawson} R.~I.,  {Lee} E.~J.,   {Chiang} E.,  2016, \mn@doi [\apj]
  {10.3847/0004-637X/822/1/54}, \href
  {http://adsabs.harvard.edu/abs/2016ApJ...822...54D} {822, 54}

\bibitem[\protect\citeauthoryear{{Laskar}}{{Laskar}}{2000}]{2000PhRvL..84.3240L}
{Laskar} J.,  2000, \mn@doi [Physical Review Letters]
  {10.1103/PhysRevLett.84.3240}, \href
  {http://adsabs.harvard.edu/abs/2000PhRvL..84.3240L} {84, 3240}

\bibitem[\protect\citeauthoryear{{Morris}, {Boley}, {Desch}  \&
  {Athanassiadou}}{{Morris} et~al.}{2012}]{2012ApJ...752...27M}
{Morris} M.~A.,  {Boley} A.~C.,  {Desch} S.~J.,   {Athanassiadou} T.,  2012,
  \mn@doi [\apj] {10.1088/0004-637X/752/1/27}, \href
  {http://adsabs.harvard.edu/abs/2012ApJ...752...27M} {752, 27}

\bibitem[\protect\citeauthoryear{{Muto}, {Takeuchi}  \& {Ida}}{{Muto}
  et~al.}{2011}]{2011ApJ...737...37M}
{Muto} T.,  {Takeuchi} T.,   {Ida} S.,  2011, \mn@doi [\apj]
  {10.1088/0004-637X/737/1/37}, \href
  {http://adsabs.harvard.edu/abs/2011ApJ...737...37M} {737, 37}

\bibitem[\protect\citeauthoryear{{Namouni}}{{Namouni}}{2010}]{2010MNRAS.401..319N}
{Namouni} F.,  2010, \mn@doi [\mnras] {10.1111/j.1365-2966.2009.15646.x}, \href
  {http://adsabs.harvard.edu/abs/2010MNRAS.401..319N} {401, 319}

\bibitem[\protect\citeauthoryear{{Ostriker}}{{Ostriker}}{1999}]{1999ApJ...513..252O}
{Ostriker} E.~C.,  1999, \mn@doi [\apj] {10.1086/306858}, \href
  {http://adsabs.harvard.edu/abs/1999ApJ...513..252O} {513, 252}

\bibitem[\protect\citeauthoryear{{Papaloizou}}{{Papaloizou}}{2002}]{pap2002}
{Papaloizou} J.~C.~B.,  2002, \mn@doi [\aap] {10.1051/0004-6361:20020490},
  \href
  {http://adsabs.harvard.edu/cgi-bin/nph-bib_query?bibcode=2002A%26A...388..615P&db_key=AST}
  {388, 615}

\bibitem[\protect\citeauthoryear{{Rein}}{{Rein}}{2012}]{2012MNRAS.422.3611R}
{Rein} H.,  2012, \mn@doi [\mnras] {10.1111/j.1365-2966.2012.20869.x}, \href
  {http://adsabs.harvard.edu/abs/2012MNRAS.422.3611R} {422, 3611}

\bibitem[\protect\citeauthoryear{{Rephaeli} \& {Salpeter}}{{Rephaeli} \&
  {Salpeter}}{1980}]{1980ApJ...240...20R}
{Rephaeli} Y.,  {Salpeter} E.~E.,  1980, \mn@doi [\apj] {10.1086/158202}, \href
  {http://adsabs.harvard.edu/abs/1980ApJ...240...20R} {240, 20}

\bibitem[\protect\citeauthoryear{{S{\'a}nchez-Salcedo} \&
  {Brandenburg}}{{S{\'a}nchez-Salcedo} \&
  {Brandenburg}}{1999}]{1999ApJ...522L..35S}
{S{\'a}nchez-Salcedo} F.~J.,  {Brandenburg} A.,  1999, \mn@doi [\apjl]
  {10.1086/312215}, \href {http://adsabs.harvard.edu/abs/1999ApJ...522L..35S}
  {522, L35}

\makeatother
\end{thebibliography}


\bsp	
\label{lastpage}
\end{document}